\documentclass[prb,twocolumn]{revtex4}
\usepackage{graphics,graphicx}
\usepackage{color}
\usepackage{wrapfig}
\usepackage{enumerate}
\usepackage{amssymb,amsmath}
\usepackage{bm}

\begin{document}
\title{Magnons in a two dimensional transverse field XXZ model}
\author{Satyaki Kar$^1$, Keola Wierschem$^2$ and Pinaki Sengupta$^2$ }
\affiliation{$^1$Theoretical Physics Department, Indian Association for
the Cultivation of Science, Jadavpur, Kolkata-700032, India.\\$^2$School of Physical and Mathematical Sciences, Nanyang Technological University, 21 Nanyang Link, Singapore 637371, Singapore.}
\date{\today}
\begin{abstract}
The XXZ model on a square lattice in the presence of a transverse magnetic field is studied within the spin wave theory to investigate the resulting canted antiferromagnet.
The small and large field regimes are probed separately both for easy-axis and easy-plane scenarios which reveal an unentangled factorized ground state at an intermediate value of the field.
Goldstone modes are obtained for the field-free $XY$ antiferromagnet as well as for the isotropic antiferromagnet with field up to its saturation value.
Moreover, for an easy-plane anisotropy, we find that {there exists a non-zero field, where magnon degeneracy appears as a result of restoration
of an U(1) sublattice symmetry and that, across that field, there occurs a magnon band crossing. For completeness, we then obtain}
%{The unentangled state appears when the boson number nonconserving term vanishes in the Holstein-Primakoff transformed Hamiltonian whereas band crossing occurs when boson hopping between the neighboring sites cease to exist. We also report the non-monotonic behavior of the staggered magnetization under the application of the transverse field.}
the system phase diagram for $S=1/2$ via large scale quantum Monte Carlo simulations using the stochastic series expansion technique.
Our numerical method is based on a quantization of spin along the direction of the applied magnetic field and does not suffer from a sign-problem, unlike comparable algorithms based on a spin quantization along the axis of anisotropy.
With this formalism, we are also able to obtain powder averages of the transverse and longitudinal magnetizations, which may be useful for understanding experimental measurements on polycrystalline samples.
\end{abstract}
\maketitle

%%%%%%%%%%%%%%%%%%%%%%%%%%%%%%%%%%%%%%%%%%%%
%
%	1. Introduction
%		a. Background and Motivation
%		b. The XXZ model and canted fields
%	2. Model
%		a. Transverse-field XXZ model
%		c. Literature review
%		d. Outline of Paper
%	2. Spin Wave Theory
%		a. Fig. 2 -- Magnon Modes
%		b. Fig. 3 -- Magnetization (SWT)
%	3. Quantum Monte Carlo
%		a. Fig. 4 -- Phase Diagram
%		b. Figs. 5 and 6 -- Powder Averages
%	5. Discussion
%		a. Gap Closings, Degenerate Modes
%		b. QMC matches SWT at the Factorization Field
%	6. Appendix
%		a. SWT Details
%		b. QMC Details (to be added)
%
%%%%%%%%%%%%%%%%%%%%%%%%%%%%%%%%%%%%%%%%%%%%

%%%%%%%%%%%%%%%%%%%%%%%%%%
% Use sections for PRB, or {\em section} for PRL  %
%%%%%%%%%%%%%%%%%%%%%%%%%%

%%%%%%%%%%%%%%%
\section{INTRODUCTION}
%%%%%%%%%%%%%%%
%{\em Introduction.---}
Quantum magnets have long served as the ideal framework
for exploring novel quantum phases and phenomena in interacting many body 
systems\cite{sub-sach}. From a theoretical standpoint, the reduced Hilbert space 
renders the systems amenable to powerful analytic and computational techniques.
Consequently, the interplay between
competing interactions, crystal electric field effects, lattice geometry and (in many
cases) geometric frustration can be studied systematically in a well-controlled 
manner. At the same time, rapid advances in material synthesis and 
characterization techniques have resulted in a wide array of quantum magnets
where many such novel quantum phases can be realized and investigated 
experimentally. Some examples include Bose Einstein condensation of magnons~\cite{bec},
spin liquid phases~\cite{spinliquids}, valence bond solids\cite{vbs,vbs2}, topologically non-trivial non-coplanar
spin textures\cite{owerre,zhou2} and magnetization plateaus\cite{mag-plat}.

The XXZ model -- and its straightforward generalizations -- remain
the standard paradigm for describing the vast majority of quantum magnets, 
making this family of hamiltonians arguably the most
intensively studied family of microscopic models of interacting many body systems.
The simple SU(2) variant of the model, in conjunction with additional terms
such as uniaxial anisotropies, on different lattice geometries yield a rich array
of field-driven phases with unique functionalities. Since many of these
novel states can be controllably realized in real quantum magnets by applying
an appropriate external magnetic field, the behavior of the XXZ model and its
multiple variants in an external field has been an active frontier of analytic
and numerical investigation. As a prototypical example, the quasi-1D compound
Cs$_2$CoCl$_4$ has been studied at length as a system that can realize an 
XXZ antiferromagnet under an applied transverse field~\cite{Breunig2013,breunig2,kenzel}. 
To date, most of the studies have utilized a
longitudinal magnetic field.~\cite{Cuccoli2003,Holtschneider2005,yunoki} In contrast, the study of a transverse field remains
relatively less studied.~\cite{Roscilde200x,Jensen2006}
%Studies of the effects of a canted magnetic field are even rarer.  
%From a theoretical standpoint, it can be argued that such a field can always be
%transformed to either longitudinal or transverse field via a simple
%rotation of the field direction, thus making the study of a canted field irrelevant.
%\red{(Keola: Is this correct? Maybe change sublattice to lattice?)} 
However, such an
investigation is important from an experimental standpoint. Often, the chemical
composition of spin compounds make it very difficult to synthesize single crystals,
and the experimental characterization has to rely on powder samples. This is 
particularly true for neutron scattering studies (both elastic and inelastic) -- possibly 
the most powerful experimental probes to identify different magnetic states. Neutron
scattering experiments require relatively large samples and for materials
where large single crystals are unachievable, one works with pellets of powder
samples which are comprised of microscopic domains of single crystals with randomly 
oriented axes. When such a sample is placed in a magnetic field, each domain
experiences a field in a different direction relative to its crystal axis and the 
measurements yield the average of fields along different directions. For a direct
comparison of theoretical studies with such experiments,  a
detailed study of the effects of a {transverse field on a XXZ model is 
important,and can be combined with results for a longitudinal field to 
estimate (approximately) the powder average.}

Aside from quantum magnets, the study of XXZ model in a transverse field is
important from quantum computational point of view as well~\cite{langari}. While a 
longitudinal magnetic field renders the model exactly solvable 
in one dimension by the Bethe ansatz, integrability is lost in the presence 
of a transverse magnetic field~\cite{mahdavifar}.
Quantum correlations give rise to entanglement, and the ability to control 
the amount of entanglement in a system by using a non-commuting field 
may play an important role in quantum technology applications~\cite{amico}.
Further, by tuning the transverse field in a XXZ model, it is possible to 
obtain an unentangled state~\cite{kurmann}.
This phenomenon of ground state factorization indicates an entanglement phase transition 
which has no classical analogue~\cite{amico2}.

Though a transverse field XXZ (TF-XXZ) model has been studied
 previously,~\cite{Roscilde200x,Jensen2006} a rigorous investigation of the 
sublattice structures as well as the magnon modes
as a function of the transverse field has been long due. In order to bridge that gap 
in the literature, in this letter we use spin wave theory (SWT) to explore the evolution 
of the magnetic ground states and their low-lying excitations as the transverse field 
strength is gradually increased. {Hamiltonian symmetries and their symmetry breakings, as well as the corresponding
degeneracies and Goldstone excitations are analyzed in detail.} We also identify the special entanglement free point in 
the phase space that appears at the so-called
factorizing field~\cite{langari} $h=h_f$. Magnon modes are obtained in the resulting 
canted AFM and magnetization along the field direction is observed. The analytical studies
are complemented by large scale quantum Monte Carlo (QMC) study using the 
stochastic series expansion technique in order to obtain the system phase diagram.
The 2D TF-XXZ model has been studied using quantum Monte Carlo method
 before~\cite{Roscilde200x}, and here our approach is essentially the same.In addition to 
identifying the different ground state phases as the parameters are varied, we extract 
powder-averaged values for the magnetization (weighted averages over the longitudinal 
and transverse field components of the magnetization), which are useful for analyzing 
the results of experimental measurements on polycrystalline samples\cite{Brambleby2015}.

\section{Model}
%{\em Model.---}
 We investigate the $S=1/2$ XXZ model with both Ising and $XY$
anisotropies in longitudinal as well as transverse external magnetic fields.
% Generic Spin Hamiltonian
A generic XYZ model in a magnetic field ${\vec h}$ can be written as 
\begin{equation}
{\cal H}=\sum_{\left<ij\right>}J_{x}S_{i}^{x}S_{j}^{x}+J_{y}S_{i}^{y}S_{j}^{y}+J_{z}S_{i}^{z}S_{j}^{z}-\sum_{i}{\vec h}\cdot{\vec S}_{i},
\end{equation}
where $J_{x},~J_{y},~J_{z}$ denote the spin exchange interactions along the $x,~y,~z$ spin axes and are summed over nearest neighbor pairs on the square lattice.
From here, a TF-XXZ model may be derived by setting the spin exchange interactions to $J_x=J_y=J_{\perp}$ with $J_z/J_{\perp}=\Delta$, 
and applying the transverse field along the $x$ axis, ${\vec h}=h{\hat x}$.
In zero field, the XXZ model is gapless for $-1\le\Delta\le1$ while gapped with an Ising anisotropy for $\Delta>1$. N\'{e}el long range order is observed 
in the gapped Ising-like phase, while the gapless $XY$-anisotropic regime also exhibits long-range N\'{e}el order but is instead characterized
by the presence of Goldstone modes due to the breaking of a continuous U(1) symmetry.

%%%%%%%%%%%%%%%%%%%%%%%%%%%%%%%%
% FIGURE 1: canting angle under an applied transverse field  %
%%%%%%%%%%%%%%%%%%%%%%%%%%%%%%%%

\begin{figure}[t]
\centering
\includegraphics[width=.92\linewidth,height=1.2in]{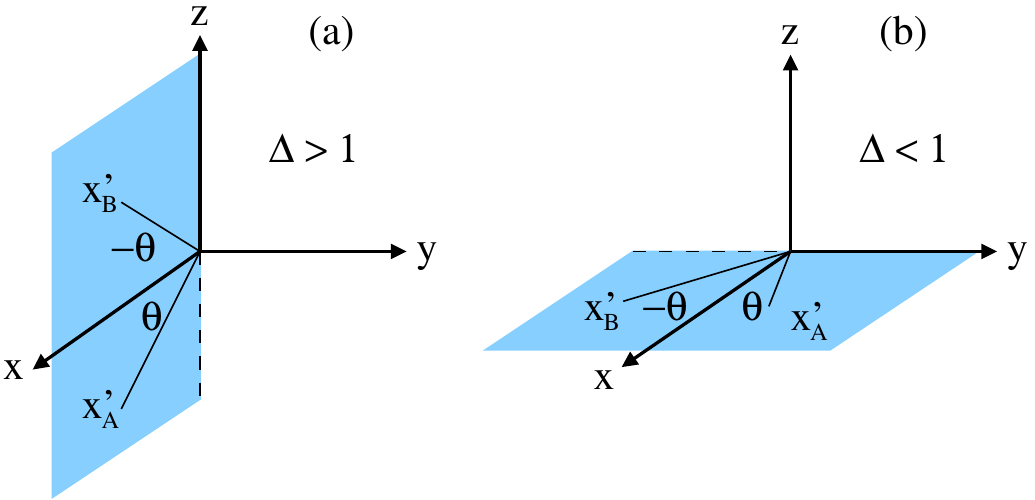}
\caption{(Color online) {Canting of the spin quantization axis in
a transverse field directed along $x$. For nonzero field, the spins are
canted parallel to the $x'$ axes.} The canting occurs (a) in $x-z$ plane (with $y'=y$) 
for $\Delta>1$ and (b) in  $x-y$ plane (with $z'=z$) for $\Delta<1$. Here $xyz$ denotes the original uncanted frame
while $x'_{A(B)}$ refers to the transformed $x$ axes in the A(B) sublattices of the canted frame.}
\label{canting}
\end{figure}

The U(1) symmetry of the XXZ model is lost upon adding the transverse 
magnetic field. At zero magnetic field, there is no magnetization in the system and the quantization axis is
decided by the exchange anisotropy parameter $J_z/J_{\perp}=\Delta$ yielding an easy-axis antiferromagnet (AFM) for $\Delta>1$
and an easy-plane AFM for $\Delta<1$. Magnetic field turns on the magnetization in the system.
With a transverse field along $x$ direction, total spin along exchange anisotropy direction becomes non-conserving away from the Heisenberg point $\Delta=1$.
A perpendicular AFM order appears with spins canted towards the field direction.
In other words, we obtain simultaneous spin alignment along the $x$ direction and AFM ordering in the $z$ (for $\Delta>1$)
or $y$ (for $\Delta<1$) direction (see Fig.~\ref{canting}). The magnetization along $x$ direction ($m_x$)
increases monotonically with magnetic field $h$ until it reaches the critical field $h = h_c$ where AFM order is extinguished and spins
align almost completely (for $\Delta \ne 1$) in the $x$ direction, forming a (nearly) saturated paramagnetic phase. However, it needs an infinitely large field, away from the Heisenberg point,
to ensure complete polarization along the field.

% Literature Review
In addition to describing the quasi-one-dimensional magnet Cs$_2$CoCl$_4$ for fields applied along the 
$b$-axis~\cite{Breunig2013}, the TF-XXZ model is also related to effective models
for certain quantum magnets where an alternating $g$-tensor and/or Dzyaloshinskii-Moriya interaction can give 
rise to an effective staggered field~\cite{Oshikawa1997}. Such an effective model has successfully been applied to
the quasi-one-dimensional quantum magnet copper benzoate~\cite{Oshikawa1997,Essler1999,Zvyagin2011}.

In this work, we focus on the case of a uniform magnetic field perpendicular to the axis of exchange anisotropy in a spin-1/2 
XXZ model on the square lattice. The easy-axis version of this model has previously been considered by 
Jensen {\it et al.}~\cite{Jensen2006} using a Green's function approach. Their main conclusion was that for small fields, 
the reduction in spin fluctuations dominates over the spin canting, leading to an increase in the staggered magnetization 
$m_s$ along the Ising axis, as well as to an increase in the N\'eel temperature $T_N$. At higher fields, of course, the 
trend reverses, until both $m_s$ and $T_N$ are zero at the {critical} field.

% Outline of Paper

%%%%%%%%%%%%%%%%%
\section{SPIN WAVE THEORY}
%%%%%%%%%%%%%%%%%

%{\em Spin Wave Theory.---}
In order to develop the spin wave analysis for the transverse field XXZ model with magnetic field $h$ along $x$ direction, we need to first 
identify how the quantization direction changes with $h$. An Ising anisotropy causes the spin quantizations
in the two sublattices to be along $\pm z$ directions. But
U(1) symmetry in the XY anisotropic case forbids any such preferences for quantization direction in the $xy$ plane. With infinitesimal $h$ along $x$, however, 
the symmetry is broken and spin flop process results in the perpendicular $\pm y$ directions to stand out as the quantized axes (see Fig.~\ref{canting}). 
As $h$ is increased, the sublattice magnetization starts canting towards $x$ direction until it becomes parallel to $x$
axis, though 
{the maximal value of the spin is reached at an infinite value of $h$ in presence of exchange anisotropy. 
There exists a finite critical value of the field at which the spins align
parallel to the field -- this is marked by  a sharp 
change in the slope of the $m_x$ vs. $h$ curve with the magnetization close to its 
saturation value. Beyond this critical field, the magnetization increases slowly (due to 
decrease in quantum fluctuations)
towards full polarization which is reached theoretically at an infinite field}
%{Actually magnetization saturates almost to maximum at the finite critical field. From
%that point on, magnetization very slowly increases (due to decrease of quantum fluctuations)
%towards the full polarization to reach there theoretically at an infinite field.} 
At some non-zero $h=h_f$, a factorized ground state is obtained where entanglement 
becomes zero. 
%Considering that state to be the spin reference state we obtain the spin wave modes of the system.In order to achieve that, 
In the case of an Ising (XY) anisotropy, we first perform a spin-coordinate rotation by an angle $\pm\theta$ about the 
spin-$y$ (spin-$z$) axis in the A ($\uparrow$) and B ($\downarrow$) sublattices respectively.
Calling the canted new $x$ directions to be the quantization directions, a ferromagnetic state is obtained in the transformed coordinates.

Within the linear spin wave approximation in this rotated frame, the easy-plane XXZ Hamiltonian gets transformed to 
(for the remainder of this section we set $J_{\perp}$ to unity and use it as our unit of measurement),
\begin{align}
H&=E_0(\theta)+\sum_{<ij>}[\frac{h}{Z}(n_i+n_j)cos\theta-cos(2\theta) S(n_i+n_j)\nonumber\\
&+\frac{{\rm cos}2\theta-\Delta}{4}(a_i^\dagger b_j^\dagger+hc)+\frac{cos~2\theta+\Delta}{4}(a_i^\dagger b_j+ hc)\nonumber\\
&+\frac{2h{\rm sin\theta/Z-sin}2\theta}{4}(a_i^\dagger-b_j^\dagger+{\rm hc})] .\nonumber
\end{align}
Here $n_i~(n_j)$ and $a_i~(b_j)$ are spin deviation and bosonic annihilation operators respectively at site $i~(j)$ within the $\uparrow~(\downarrow)$ 
sublattice and $Z$ is the coordination number ($Z=4$ in 2D XXZ model).
See {appendix \ref{ap1}} for details.
Minimizing $E_0(\theta)$ identifies the state of quantization by selecting the reference angle $\theta_r$ with cos$\theta_r=h/2ZS$.
%Thus we fix the reference state for spin wave expansion.
A Fourier transformation, from there on, leads to
\begin{eqnarray}
H&=&E_0(\theta_r)+\sum_{k}[ZS(a_k^\dagger a_k+b_k^\dagger b_k)+ZS\gamma_k(\frac{{\rm cos}2\theta_r+\Delta}{2}\nonumber\\
&&(a_k^\dagger b_{k}+hc)+\frac{{\rm cos}2\theta_r-\Delta}{2}(a_k^\dagger b_{-k}^\dagger+hc))]
\label{ham-f0}
\end{eqnarray}
with {$\gamma_k=(cos~k_x+cos~k_y)/2$.}
We need to resort to a $4\times4$ Hamiltonian matrix formulation~\cite{mano} to solve this problem {(see appendix \ref{ap2})}.
A Bogoliubov transformation for such case~\cite{kar} brings in the magnon modes to be given by 
$\Omega_k=\sqrt{(ZS\pm\frac{ZS\gamma_k({\rm cos}2\theta_r+\Delta)}{2})^2-(\frac{ZS\gamma_k({\rm cos}2\theta_r-\Delta)}{2})^2}$.
{The easy-planar AF, for $h=0$, has no preferred quantization directions in the
$xy$-plane and hence enjoys a U(1) symmetry. Switching on the field, even infinitesimally,
spontaneously breaks that symmetry causing non-degenerate acoustic (with Goldstone excitation) and optical magnon modes to appear. Gradual increase in $h$
reduces the gap between the modes, eventually restoring magnon degeneracy at $h=h_d=2SZ\sqrt{(1-\Delta)/2}$. At this point the Holstein-Primakoff transformed Hamiltonian lacks the boson hopping term between neighboring sites.
We will see that for easy axis or isotropic case, such vanishing of the hopping term occurs at $h=0$ and magnon modes become degenerate there as well}.
Now also notice that for $cos(2\theta_r)=\Delta$, Eq.~\ref{ham-f0} is devoid of the number-nonconserving third term and the spin reference state
indeed becomes the ground state. Hence we realize a
factorized ground state which indicates zero quantum entanglement. This is parametrized as
cos$\theta_f=\sqrt{\frac{1+\Delta}{2}}$ and $h_f=2ZS$cos$\theta_f$.

Similarly for the easy axis scenario, we obtain
%{
\begin{align}
H&=E_0(\theta_r)+ZS\sum_{k}[\Delta (a_k^\dagger a_k+b_k^\dagger b_k)+\gamma_k(\frac{{\rm cos}^2\theta_r(1+\Delta)}{2}\nonumber\\
&(a_k^\dagger b_{k}+hc)+\frac{2-{\rm cos}^2\theta_r(1+\Delta)}{2}(a_k^\dagger b_{-k}^\dagger+hc))].
\label{ham-f2}
\end{align}
%}
with  $cos\theta_r=h/SZ(1+\Delta)$. The factorizing point is denoted by  cos$\theta_f=\sqrt{\frac{2}{1+\Delta}}$ and $h_f=ZS(1+\Delta)$cos$\theta_f$.
The magnon modes are given by 
$\Omega_k=\sqrt{(\Delta SZ\pm\frac{ZS\gamma_k{\rm cos}^2\theta_r(1+\Delta)}{2})^2-(\frac{ZS\gamma_k(2-{\rm cos}^2\theta_r(1+\Delta))}{2})^2}$.

Hence, with the application of a transverse field $h$, the degeneracy between the magnon modes within the reduced Brillouin zone is lost. 
Even at $h=0$, non-degenerate modes are obtained as long as  $\Delta<1$. Goldstone modes are present for all values of $XY$ anisotropy whereas the system exhibits a finite gap to lowest magnetic excitations for $\Delta>1$.
%%%%%%%%%%%%%%%%%%%%%%%%%%%%%%%%%%%%%%%%
% FIGURE 2: evolution of magnon modes from SWT under a transverse field  %
%%%%%%%%%%%%%%%%%%%%%%%%%%%%%%%%%%%%%%%%
\begin{figure}[t]
\centering
\includegraphics[width=0.9\linewidth,height=2.2 in]{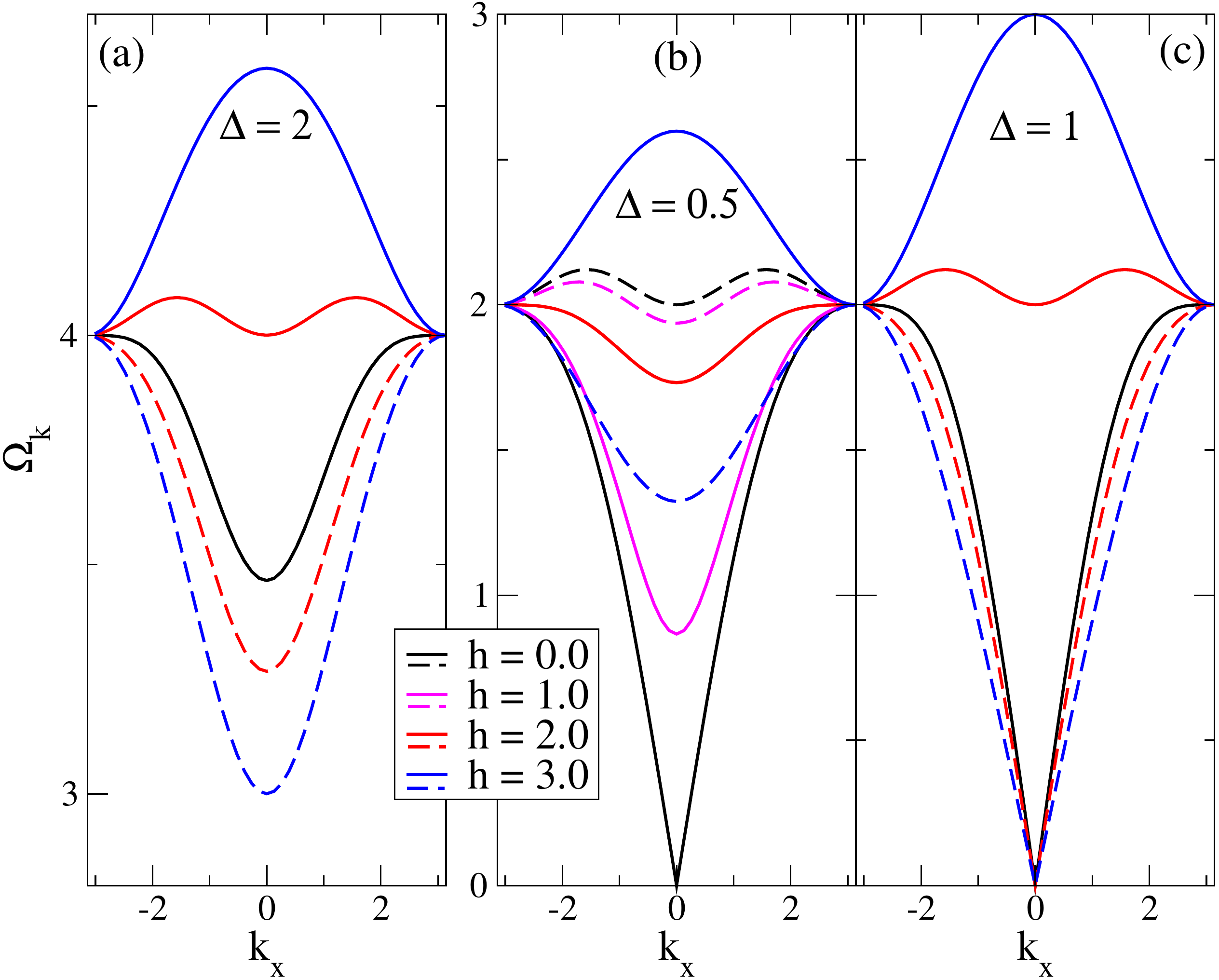}
\caption{(Color online) Magnon modes $\Omega_k$ of the $S=1/2$ XXZ model on the square lattice as a function of momentum $k_x$ 
(with $k_y=0$) for various values of the transverse field $h$ and spin exchange anisotropy (a) $\Delta=2.0$ and (b) $\Delta=0.5$ and (c) $\Delta=1.0$.}
\label{disp}
\end{figure}

{We can understand the behavior of the magnon excitation modes intuitively from symmetry considerations. Let us first discuss the
field-free XXZ model at $h=0$. For $\Delta>1$, spin quantization directions are along $z$. 
There is a $Z_2$ symmetry corresponding to the transformation $S_{i,z}\rightarrow-S_{i,z}$ (denoted by $Z_{2,z}$) as well as an $U(1)$ symmetry
corresponding to $(S_{i,x}+iS_{i,y})=S_i^+\rightarrow S_i^+e^{i\phi}$ (denoted by $U_{xy}(1)$) for arbitrary angle $\phi$ about $z$. Thus the Hamiltonian possess an overall 
$ Z_{2,z}\bigotimes U_{xy}(1)$ symmetry and, consequently, two degenerate   magnon modes. This remains true up to the isotropic limit when
an overall $SU(2)$ symmetry is observed in the Hamiltonian. Now a Goldstone excitation results if a continuous symmetry of the Hamiltonian is broken
spontaneously by the ground state in the thermodynamic limit. The spin component along the quantization direction is a good
quantum number and for $\Delta=1$, this can be continually rotated leaving the 
Hamiltonian intact and thereby yielding Goldstone modes in the spectrum. For $\Delta>1$, the quantized component $S_z$ does not have that liberty due to spin 
anisotropy and no Goldstone excitation is formed. 
For $\Delta<1$, the quantization direction changes (see Fig.\ref{canting}).
Considering this direction to be along $y$ (which will be the case due to spin-flopping, with a transverse field along $x$ direction), 
we see that the Hamiltonian still possesses an $U_{xy}(1)$ symmetry enabling the system to have a Goldstone mode (however, note that, a $U_{yz}(1)$ 
symmetry is not obeyed and hence only one Goldstone mode is observed in this case).
The discrete $Z_{2,y}$ symmetry is obeyed.  However, $U_{xz}(1)$ symmetry is lost because of the spin-anisotropy of the rest of
the terms: $J(S_{i,x}S_{j,x}+\Delta S_{i,z}S_{j,z})$. This, in turn, makes the magnon modes nondegenerate.
}

{Switching on a non-commuting transverse field $h$ results in an interesting outcome. Spin canting develops and the quantization 
directions (denoted by $x'_A$ and $x'_B$ for A and B sublattices) in the two sublattices no longer remain oppositely 
directed. A magnon degeneracy, in this case, would require an $U_{y'z'}(1)$ symmetry corresponding to sublattice rotations about $x'$ 
axes (by angle $\phi$ and -$\phi$, say, for the two sublattices respectively). But that is absent as canting causes other 
phase-nonconserving terms to appear in presence of $h$. So degeneracy is lifted, in general.
% For $\Delta>1~(\Delta<1)$, a magnon degeneracy would require an 
% $U_{y'z'}(1)~(U_{x'y'}(1))$ symmetry which is lost due to canting (remember that by primed frames,
% we denote the rotated axes), because in presence of $h$. Hence the magnon modes become nondegenerate. 
% This holds for $\Delta=1$ as well.
The isotropic point at $\Delta=1$, however, holds a sublatttice symmetry corresponding to a continuous rotation of $S_x'$
by any angle $\phi$ (and $-\phi$ on the other sublattice) about an axis which lies in the $x'x$ plane. This results in a Goldstone mode (and not
two Goldstone modes because of the restriction on the axis of rotation) appear
which survives till $h<h_c$. Beyond $h_c$, both $x'_A$ and $x'_B$ overlap with the $x$ direction ruling out any spontaneously broken 
symmetry for the ground state. Next, we see that for $h=h_d$,
magnon degeneracy resurfaces for planar anisotropy. This is the singular point where coefficients of the fluctuation terms $S_i^+S_j^-$
vanish and the Hamiltonian is invariant under a sublattice rotation by an arbitrary angle $\phi$ (and -$\phi$ on the other 
sublattice) in the $y'-z'$ plane.
%transforming $(S_{y'}+iS_{z'})=S'^+\rightarrow S'^+e^{i\phi}$. 
This reappearance of $U_{y'z'}(1)$ sublattice symmetry brings back degenerate magnon modes.}

{For $\Delta<1$, the lower and higher magnon branches start moving towards each other as $h$ is increased from zero
and eventually a magnon band crossing occurs at $h=h_d$. At the critical field $h_c$, the magnon spectrum contains an acoustic and 
an optical mode. In contrast, an isotropic AF has degenerate acoustic modes at $h=0$ whereas easy-axis AF has degenerate
optical modes (i.e., the minimum magnon energy is positive). In either case a finite $h$ lifts
the magnon degeneracy resulting in the appearance of acoustic-optical mode pair at $h=h_c$. 
But there is no band crossing.
However, for easy-plane AF, we see an acoustic and an optical mode due to spontaneous breaking of the
U(1) symmetry at $h=0$. With increase in $h$, gap between the modes
reduces, they cross each other at some intermediate field finally to become an acoustic-optical
mode pair again for $h=h_c$ (But this time, the acoustic mode at $h=0$ evolves to
become an optical mode at $h=h_c$ and the vice versa).
This feature can be observed in neutron scattering experiment, where density of states show
large intensities at the field where magnon degeneracy appear. Also, by 
experimentally probing the lowest energy excitations, a change in the
excitation spectrum  can be detected
during field tuning across the particular field exhibiting degeneracy.}

Fig.~\ref{disp} demonstrates such behavior showing the magnon dispersion plots for easy-axis $\Delta=2$, easy-planar $\Delta=0.5$ and 
isotropic $\Delta=1$
at $k_y=0$. The gapped and gapless nature of the Ising and $XY$ anisotropy respectively can be readily seen there.
%At $h=0$, the magnon modes are gapped and degenerate for $\Delta>1$. But due to U(1) symmetry breaking, Goldstone modes appear for the XY-anisotropy
%rendering the system gapless with non-degenerate modes.
%It takes some nonzero $h$ value to bring in the degeneracy in that case.

This SWT analysis (call it SWT$_a^{(1)}$) indicates a maximum field value {$h=h_c$} with {$h_c=2ZS$} for $|\Delta|<1$ and 
{$h_c=ZS(1+\Delta)$} for $\Delta\ge1$, up to which this method can be meaningfully
pursued as $\theta$ no more remains real beyond that. With an increase in $h$,
calculations eventually lead to a gapless mode at {$h=h_c$} thereby indicating
the limit beyond which the choice of given reference state fails. 
In this regard, it may also be pointed out that with an increase in $h$, the spin deviation $\epsilon$ (a measure of quantum fluctuation
in this case) consistently decreases to become zero at $h=h_f$.

For large $h$ ($i.e.,~h\ge {h_c}$), $E_0(\theta)$ also become minimum at $\theta=0$ and we consider, instead, a different prescription (call it SWT$_b^{(1)}$), with
the ferromagnetic state along $x$ direction being the new spin reference state. The SWT$_b^{(1)}$ calculation gives the magnon dispersion expression to be
$\Omega_k=\sqrt{(h-SZ+\frac{1+\Delta}{2}\gamma_kSZ)^2-(\frac{\Delta-1}{2}\gamma_kSZ)^2}$ where the measure of the critical field for full
polarization becomes ${h_c'}=SZ+\frac{1+\Delta}{2}SZ+\frac{|\Delta-1|}{2}SZ$ {(see appendix \ref{ap3})}. Notice that $h_c={h_c'}$, as it should be. Additionally at the Heisenberg point,
we obtain $h_c=h_f$ as well.
%%%%%%%%%%%%%%%%%%%%%%%%%%%%%%%%%%%%%%%
% FIGURE 3: comparison of magnetization curves between QMC and SWT %
%%%%%%%%%%%%%%%%%%%%%%%%%%%%%%%%%%%%%%%

\begin{figure}[t]
\centering
\includegraphics[width=0.9\linewidth,height=2.4 in]{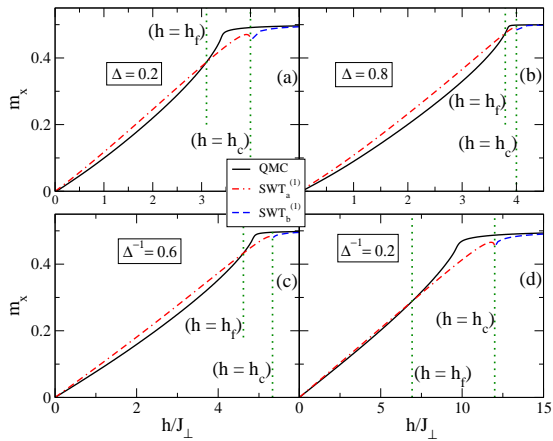}
\caption{(Color online) Magnetization $m_x$ along the field direction of the $S=1/2$ XXZ model on a square lattice 
for (a) $\Delta=0.2$, (b) $\Delta=0.8$, (c) $\Delta^{-1}=0.6$ and (d) $\Delta^{-1}=0.2$,
as a function of the transverse field $h$. The comparison between QMC and SWT$_a^{(1)}$, SWT$_b^{(1)}$ results are given.
The locations of $h_f$ and $h_c$ are also shown by the dotted lines.}
\label{Ms}
\end{figure}

The sublattice magnetization along the rotated $x$ directions can be obtained as $m_s=S-\epsilon$ where spin deviation
$\epsilon=\frac{1}{N}\sum_k(<a_k^\dagger a_k>+<b_k^\dagger b_k>)$. From there the magnetization along the field direction 
can be obtained as $m_x=m_s$cos$\theta_r$. The plots of $m_x$ for various $h$ are shown in Fig.~\ref{Ms} highlighting also
the results from QMC calculations to be discussed below.
Notice that the magnetization as obtained by linear spin wave analysis and QMC match exactly at $h=0$ and $h_f$. At $h=0$, 
the rotated quantized directions are perpendicular to $x$ direction thereby ensuring that $m_x=0$ there. On the other hand, $h_f$ is the
factorization point where we get the factorized ground state with $\epsilon=0$ and thus magnetization becomes $m_x=S$cos$\theta_f$.
At the factorization field, the ground state is a believed to be a direct product state, which explains the agreement between QMC simulations 
and SWT$_a^{(1)}$ analysis (i.e. there are no quantum corrections at $h_f$).

\begin{figure}[t]
\centering
\includegraphics[width=0.9\linewidth,height=2.2 in]{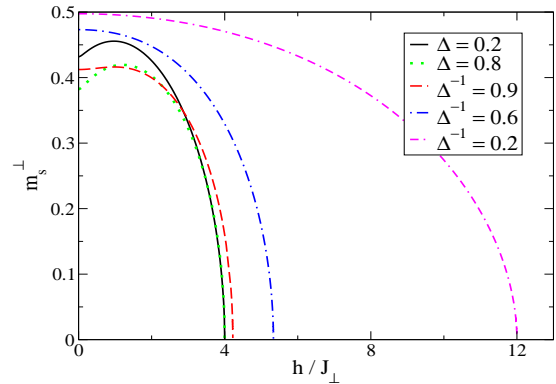}
\caption{(Color online) Staggered Magnetization $m_s^\perp$, orthogonal to the field direction (see section III for details) in a $S=1/2$ XXZ model on a square lattice, as a function of the transverse field $h$ for different values of $\Delta$.}
\label{stag}
\end{figure}

{Another quantity of interest, in this reference, is the staggered magnetization $m_s^{\perp}$ orthogonal to the field direction and along the spin quantization direction at zero field (infinitesimally small field, in the easy-plane case, however). For transverse field along $x$, these are the $z$ or $y$ directions in an easy-axis or easy-plane XXZ model respectively (see Fig. 1). Thus $m_s^{\perp}$ is obtained as $m_s^{\perp}=m_s$sin$\theta_r$. Fig. \ref{stag} shows the plot of $m_s^{\perp}$ as a function of transverse field $h$. A reduction of spin fluctuation with field (for $h<h_f$) causes $m_s^{\perp}$ to increase while a spin canting towards the field direction reduces
the magnetization component along the perpendicular direction. These two effects together determine the behavior of $m_s^{\perp}$ under the variation of the field.
In the easy-plane XXZ model, the former (latter) one dominates for small (large) field values and we see $m_s^{\perp}$ initially to increase with $h$, then to pass through a maximum, finally  to decrease down to zero at $h_c$ (see also Ref.\onlinecite{Jensen2006}). For large anisotropy ($i.e.,$ large $\Delta$), however, spin fluctuations are never strong enough to cause such initial increase in $m_s^{\perp}$ for small field values.}

We should also mention here that the spin deviation $\epsilon$ leaves room for correction to the spin wave results as that is what contributes
to the next higher order spin wave expansion.
At $h=0$, $\epsilon$ decreases as we move away from the Heisenberg point.
 But $\epsilon$ also consistently decreases with $h$ becoming zero at the factorization point. Beyond $h_f$, $\epsilon$ increases again,
more sharply for
$\Delta$ sufficiently away from unity. This indicates the fluctuations around the QCP and demonstrate the inability of mean
field SWT to describe the physics precisely. That is why, in Fig.~\ref{Ms}, the magnetization plots around $h_c$ show some unphysical 
turning, already witnessed for an Ising AFM (see Ref.~\citenum{mila}). As the Heisenberg point has $h_c=h_f$,
$\epsilon$ remains zero there and a linear SWT remains a good theory. But away from $\Delta=1$, $\epsilon$ starts
getting bigger with larger spin anisotropy making mean field SWT estimates more inappropriate at $h\sim h_c$.
So the phase boundaries obtained using linear SWT differs more from QMC estimates in Fig.\ref{fig-phase} for $\Delta$ further away from unity.
See that the unphysical behavior in $m_x$ for $h\sim h_c$ also gets pronounced
mostly away from $\Delta=1$ (compare Fig.\ref{Ms}(a),(d) results with that of Fig.\ref{Ms}(b),(c)).

A linear SWT {($i.e.,$ SWT$_a^{(1)}$ and SWT$_b^{(1)}$)}, thus,  can not predict an accurate phase boundary, as compared to the QMC calculations. However, 
we notice that a second order correction to linear SWT {(see appendix \ref{ap4})} improves the result and also give phase boundaries
close to the QMC predictions {(see SWT$^{(2)}$ results in Fig.\ref{fig-phase}(a))}. A perturbation analysis {(see appendix \ref{ap6})}
at the cross-over point between full polarization and the one with all but one spin flipped also describes 
the transitions better and give phase boundaries close to that obtained by QMC.

{
\subsection*{Quasi-1D models}
Following our calculations, magnon modes can also be obtained for quasi-1D XXZ model. For $f$ being the fraction of the spin
exchange interaction strength along the $y$ direction, as compared to that along $x$, the magnon dispersion $\Omega_k$ is
given as
\begin{align}
 &\Delta<1:~\Omega_k^2=\nonumber\\
 &((1+f)\pm\gamma'_k({\rm cos}2\theta_r+\Delta))^2-(\gamma'_k({\rm cos}2\theta_r-\Delta))^2~,\nonumber\\
 &\Delta>1:~ \Omega_k^2=\nonumber\\
 &(\Delta(1+f)\pm\gamma'_k{\rm cos}^2\theta_r(1+\Delta))^2-(\gamma'_k(2-{\rm cos}^2\theta_r(1+\Delta)))^2
\end{align}
where $\gamma'_k=[cos(k_x)+fcos(k_y)]/2$.
This is a good estimate for elementary excitations as long as
$x$ is not very small, because deconfined spinons appear otherwise
affecting the excitation modes\cite{tsvelik}.}
%\subsection{Single-ion anisotropy}
%With single-ion anisotropy, Hamiltonian becomes $H=H_0+D\sum_iS_{iz}^2$, as $\hat z$ denotes the easy direction in the XXZ model. 
%$D$ only offers an effective longitudinal field within LSW theory and
%comparing the values of $D$ and transverse field $h$, we need to decide on the suitable spin reference state.}
% \red{As we look at Eq.A4, we realize that the first term is invariant under 
% both U(1) symmetry along xy plane and so is the second term. The 3rd term has a restricted symmetry of rotation by $\pi$ about $z$ direction.}

%%%%%%%%%%%%%%%%%%%%
\section{QUANTUM MONTE CARLO}
%%%%%%%%%%%%%%%%%%%%

%{\em Quantum Monte Carlo.---}
The typical way of dealing with transverse fields within the stochastic series expansion (SSE) formalism, or QMC more generically, has been to treat them as adding individual raising and lowering operators to the XXZ Hamiltonian. This method has been successful in describing ferromagnetic systems, and details of this typical implementation of transverse fields can be found in Refs.~\citenum{Henelius200x} and~\citenum{Syljuasen2003}. However, this approach is not suitable for antiferromagnetic models, as the off-diagonal nature of the transverse field complicates the sublattice rotation necessary to transform the Hamiltonian into a sign-problem-free form.

In this work, we take an alternative approach by choosing the direction of the applied magnetic field as the projection axis for spin quantum number so that the magnetic field acts upon the spins via diagonal operators {(see appendix \ref{ap5})}. The Hamiltonian for the TF-XXZ model is given by
\begin{equation}
{\cal H}=J_{\perp}\sum_{\langle ij\rangle}S_{i}^{x}S_{j}^{x}+S_{i}^{y}S_{j}^{y}+\Delta S_{i}^{z}S_{j}^{z}-h\sum_{i}S_{i}^{x}.
\end{equation}
Choosing the $x$ axis as our spin quantization axis, we rewrite the above Hamiltonian in terms of the ladder operators $S^{\pm}=S^{y}\pm iS^{z}$ to find
\begin{eqnarray}
\begin{split}
{\cal H}=J_{\perp}\sum_{\langle ij\rangle}S_{i}^{x}S_{j}^{x}+
\frac{1-\Delta}{4}\left(S_{i}^{+}S_{j}^{+}+S_{i}^{-}S_{j}^{-}\right)+\\
\frac{1+\Delta}{4}\left(S_{i}^{+}S_{j}^{-}+S_{i}^{-}S_{j}^{+}\right)-h\sum_{i}S_{i}^{x}.
\end{split}
\end{eqnarray}
This Hamiltonian can be shown to be free of the QMC ``sign problem'' for bipartite lattices by choosing an appropriate (sub)lattice rotation, or by keeping track of the overall sign of the vertex weights in the operator string---for more details, see the appendix.

In order to accommodate planar anisotropy, additional vertices need to be included compared to the standard ones required for axially anisotropic Hamiltonians. This was noted by Roscilde {\it et al.} in their earlier study of the TF-XXZ model~\cite{Roscilde200x}, 
and has also been discussed in relation to the quantum compass model on a square lattice by Wenzel {\it et al.}~\cite{Wenzel2008,Wenzel2010}. Here, we comment that while the added terms $S_{i}^{+}S_{j}^{+}$ and $S_{i}^{-}S_{j}^{-}$ break the U(1) symmetry of the 
zero-field XXZ model, they preserve a $Z_2$ symmetry corresponding to the total magnetization modulo 2. This turns out to be sufficient to guarantee that link discontinuities in the directed loop update can only occur in pairs, and therefore we may use the 
standard directed loop equations (though they now act on 4x4 matrices of vertex weights--we use ``solution B'' of Sylju\r{a}sen~\cite{Syljuasen2003}).

%%%%%%%%%%%%%%%%%%%%%%%%%%%
% FIGURE 4: phase diagrams from QMC and SWT %
%%%%%%%%%%%%%%%%%%%%%%%%%%%

\begin{figure}
\centering
\includegraphics[width=\linewidth]{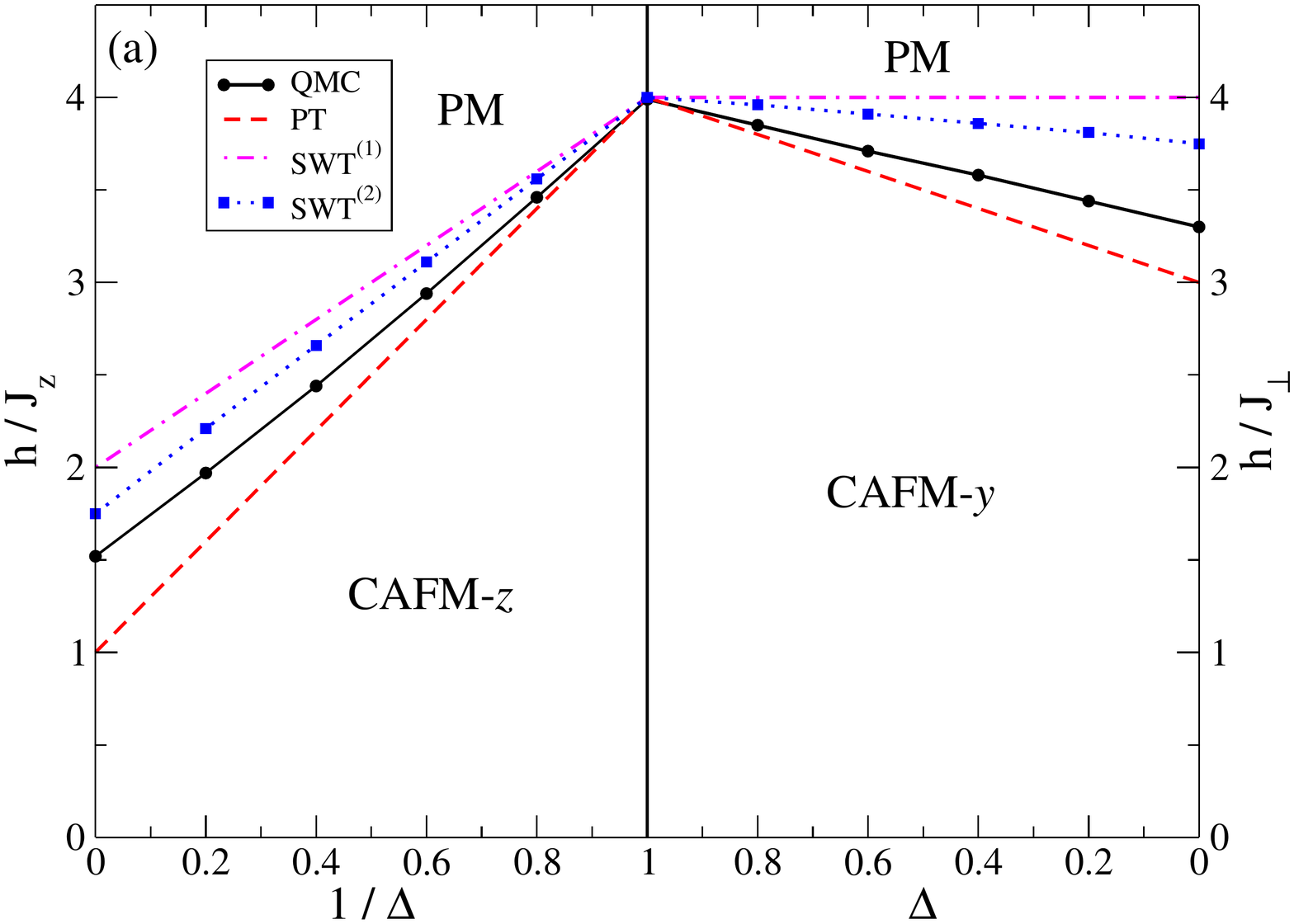}
\\
\vskip -.2 in
\includegraphics[width=\linewidth]{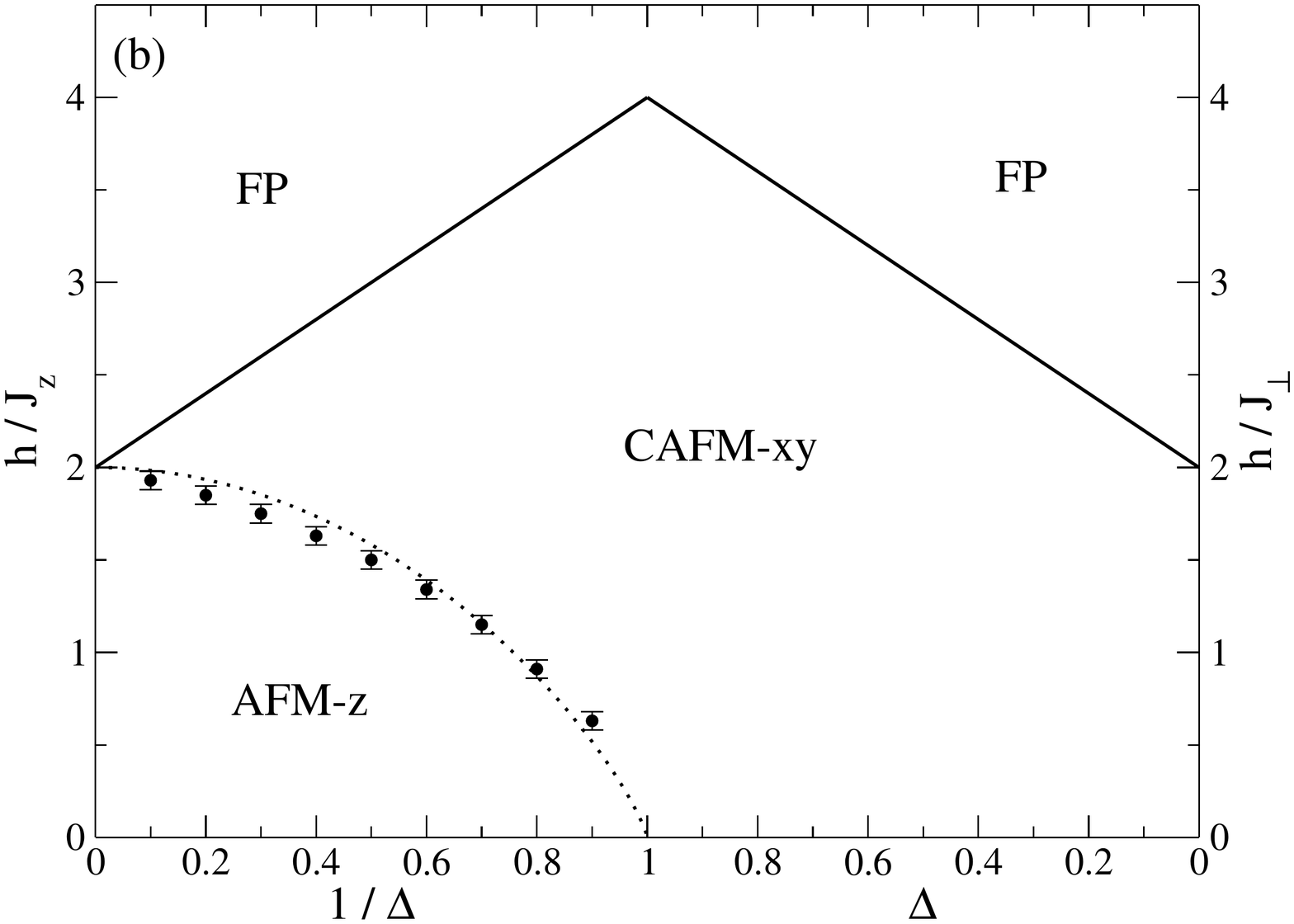}
\caption{(Color online) Phase diagram of the $S=1/2$ XXZ model in (a) transverse (with $\vec h=h\hat x$) and (b) 
longitudinal (with $\vec h=h\hat z$) magnetic fields. 
Within (a), phase boundaries between the canted Ising states (CAFM-z and CAFM-y) and the nearly saturated paramagnetic (PM) phase are shown for 
quantum Monte Carlo (QMC), first order perturbation theory (PT), first order spin wave theory (SWT$^{(1)}$), and second order 
spin wave theory (SWT$^{(2)}$). Solid black lines represent  
exact boundaries by QMC. Dashed lines represent analytic results from PT (red) and SWT$^{(1)}$ (magenta) while
numerical data points from QMC (black) and SWT$^{(2)}$ (blue) are shown as well.
{QMC data points in panels (a) and (b) are determined by finite-size crossings of $\rho_sL$ and energy level crossings, respectively.}}
\label{fig-phase}
\end{figure}

Using the QMC scheme described above, we have obtained the magnetic phase diagram as a function of spin 
exchange anisotropy and applied magnetic field (fig.\ref{fig-phase}). 
For a field along the longitudinal direction, $h_z$, the ground state phase diagram is relatively
simple and well-known. At the isotropic point ($\Delta = 1$), in the absence of any external
field the system is in a gapless N{\' e}el phase  with a spontaneously chosen quantization axis. 
When a field is turned on, the AFM ordering is confined to the $xy$ plane,
and a non-zero uniform magnetization is induced parallel to the applied field. We refer to
this canted AFM phase as CAFM-$xy$. The canting increases monotonically with increasing
field and the system becomes fully polarized at a saturation field,  {$h_{s}=ZS(J_{\perp}+J_{z})$}.
Interestingly, the expression for the saturation field is an exact result. Away from the 
Heisenberg point, for $XY$-like anisotropy ($\Delta < 1$), the ground state at zero field has long 
range AFM order with spontaneously broken symmetry in the $xy$ plane (AFM-$xy$) and 
gapless excitations.  The field
induced behavior is qualitatively similar to that in the Heisenberg limit -- the ground state
acquires a canting of the spins parallel to the field (CAFM-$xy$) which increases monotonically
up to saturation. For Ising-like anisotropy ($\Delta > 1$), the ground state is characterized by 
longitudinal AFM order with a finite gap to lowest spin excitations. With increasing field, the 
system remains in the AFM-$z$ phase up to a critical point, at which point there is a transition to 
the CAFM-$xy$ phase accompanied by the closing of the spin gap. {The critical field of this first-order phase 
transition can be determined by an energy level crossing in the QMC data.} Upon further increasing the 
field, the canting increases till it reaches saturation. 

The situation is more complex for transverse field. As shown in Fig.~\ref{fig-phase}, under a small 
transverse field the XXZ model displays two phases: canted AFM-$y$ phase (CAFM-$y$) and the canted 
AFM-$z$ phase (CAFM-$z$). The CAFM-$y$ and CAFM-$z$ phases possess uniform magnetization along the 
$x$-axis simultaneously with antiferromagnetic order along the $y$- and $z$-axes, respectively. The 
canting along the respective axes increase monotonically, but the system reaches saturation only at an 
infinite field strength. Instead, there is a critical field above which long range order is lost and 
the system enters a partially polarized state. Up to first order in perturbation theory {(see appendix \ref{ap6})} the critical
field is estimated as {$h_{c}=ZS(3J_{\perp}+J_{z})/2$}.
The phase transition at this critical field is continuous and belongs to the Ising universality class in $2+1$ 
dimensions\cite{pelissetto}. This field can be accurately determined with QMC data by using finite-size scaling of the structure 
factor of the staggered magnetization along the $y$ or $z$ axis.
{It can be pointed out here that the mean field SWT overestimates the critical field and thus it is, in general, higher than the 
values obtained using QMC.}

%%%%%%%%%%%%%%%%%%%%%%%%%%%
% FIGURE 5: powder averages for XY anisotropies %
%%%%%%%%%%%%%%%%%%%%%%%%%%%

\begin{figure}
\centering
\includegraphics[width=0.98\linewidth,height=2.6 in]{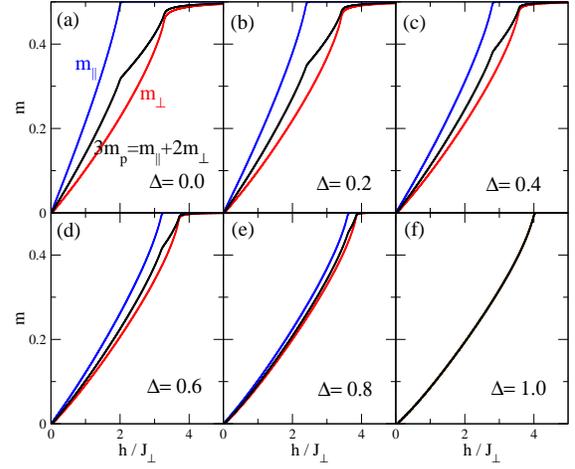}
\caption{(Color online) Powder averages for $XY$-like anisotropies. Transverse field data in red, longitudinal field data in 
blue, and powder average in black.}
\label{fig-powder-xy}
\end{figure}
%%%%%%%%%%%%%%%%%%%%%%%%%%%%
% FIGURE 6: powder averages for Ising anisotropies  %
%%%%%%%%%%%%%%%%%%%%%%%%%%%%

\begin{figure}[htp]
\vskip .1in
\centering
\includegraphics[width=0.98\linewidth,height=2.6 in]{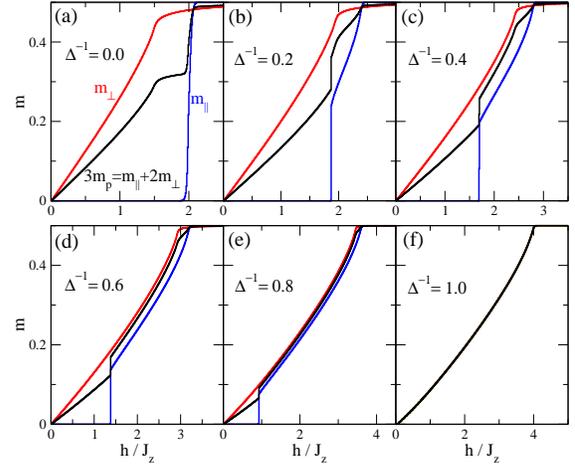}
\caption{(Color online) Powder averages for Ising-like anisotropies. Transverse field data in red, longitudinal field data in 
blue, and powder average in black.}
%\red{Keola says: try comparing to the magnetization curves of polycrystalline organometallic antiferromagnets as shown in Fig.~10 of Brambleby {\it et al.}~\cite{Brambleby2015}.}
\label{fig-powder-z}
\end{figure}
The powder average for magnetization is given by 
{
\begin{equation}
3m_{p}=2m_{\perp}+m_{||}
\label{pow}
\end{equation}
where $m_\perp$ and $m_{||}$ are magnetizations for external magnetic field 
perpendicular and parallel to the easy direction $(i.e.,~z)$ respectively. So $m_{||}=m_z$ for longitudinal fields along the $z$ direction and
$m_\perp=m_x$ for transverse field along the $x$ direction.
Within QMC, these are calculated as $m_{x(z)}=\frac{1}{N}\sum_iS_{i,x(z)}$. Eq. \ref{pow}} can be obtained by integrating the well-known powder
average formula for susceptibility~\cite{Rigol2007} 
%\red{(Keola says: there must be a standard reference for powder averaging--let's find it!)}.
In Figs.~\ref{fig-powder-xy} and ~\ref{fig-powder-z} we show the powder averaged magnetization $(m_p$) as a function of applied magnetic field for 
$XY$- and Ising-like anisotropy, respectively.
\begin{figure}[t]
\vskip .1in
\centering
\includegraphics[width=0.98\linewidth]{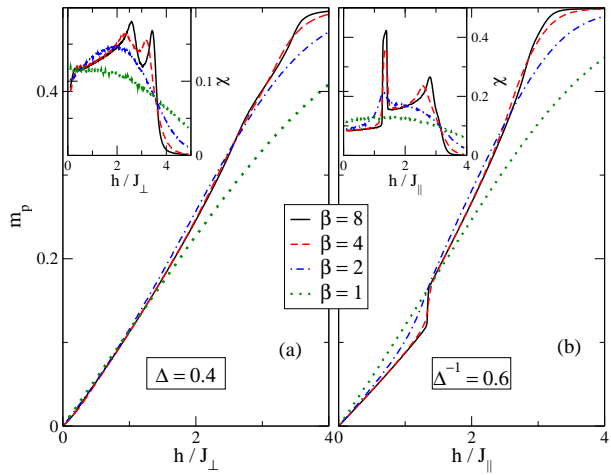}
\caption{(Color online) Powder averages $m_p$ vs. field $h$ for (a) easy-planar $\Delta=0.4$ and (b) easy-axis $\Delta^{-1}=0.6$ cases at
different temperatures. Here $\beta=1/k_BT$ values are in units of (a) $J_\perp$ or (b) $J_{||}$. The inset shows the corresponding 
susceptibilities.}
%\red{Keola says: try comparing to the magnetization curves of polycrystalline organometallic antiferromagnets as shown in Fig.~10 of Brambleby {\it et al.}~\cite{Brambleby2015}.}
\label{powder-T}
\end{figure}
{Notice that the variation of $m_p$ with field develops a kink (or jump) before the saturated field value for easy-planar (easy-axis) 
anisotropy when $\Delta$ is away from unity.
This is also realized in magnetization measurements from powder samples with easy planar anisotropy\cite{Brambleby2015}. 
Furthermore, we find that a temperature variation of 
powdered magnetization profile shows a gradual thermal smoothening of such kink-features (see Fig.\ref{powder-T}), in tune with the observations
from polycrystalline materials\cite{Brambleby2015}. The inset in Fig.\ref{powder-T} captures the behavior of the susceptibility
$\chi=dm_p/d h$ where two peaks can be witnessed at low temperatures. These peaks are due to critical points where antiferromagnetic order ceases: the first in response to the longitudinal component of the field and the second in response to the transverse component of field. For easy-planar anisotropy, peaks of
comparable height are obtained as also observed experimentally and reported in Ref.\onlinecite{Brambleby2015}. For easy axis anisotropy, on the other hand,
the first peak is a sharp one due to the sudden increase in magnetization occuring at the spin-flop transition for longitudinal component of the field.
%the second peak is at the critical point while the first one occurs at an intermediate field below $h_c$.
All these observations indicate that an analytic calculation followed by numerical computations of thermalized states 
in presence of longitudinal and transverse field contributes significantly in understanding the magnetic response from powder samples.}
\\

%%%%%%%%%%%%%
\section{DISCUSSION}
%%%%%%%%%%%%%

%{\em Discussion.---}
We have studied the ground state phases and low lying excitations of the two dimensional XXZ
model -- with both Ising-like and $XY$-like exchange anisotropies -- in the presence of a 
transverse magnetic field. The transverse field causes a tilting of the sublattice magnetization 
producing canted magnetic orders in the system. For a small field, the overall magnetization 
grows slowly as the field competes with the spin anisotropy. Both the longitudinal and
transverse components of the magnetization are probed, as is the low-lying excitation spectrum.
The evolution of the magnon 
excitation with increasing field is examined in detail using spin wave theory, with particular emphasis on the nature of the 
excitation spectrum at the entanglement free point. For this 
critical value of the field, quantum fluctuations are suppressed, resulting in an unentangled ground state at a finite field 
value. Beyond this point, however, fluctuation shoots up fast to become maximum at the transition point $h_c$. The Spin wave 
results are complemented by and benchmarked against large scale QMC simulations, yielding a deeper understanding of the 
magnetic properties across a wide range of Hamiltonian parameters. 
 We find that in a longitudinal field, the saturation field can be calculated exactly to be $h_{s}=ZS(J_{\perp}+J_{z})$. In a
transverse field, on  the contrary,
the expression is no longer exact, in part because the saturation field is replaced by a critical field. Up to first order 
in perturbation theory the critical field is given by $h_{c}=ZS(3J_{\perp}+J_{z})/2$. {We also provide an estimate of magnon excitation
modes in quasi-1D antiferromagnets.}
Finally, we use our QMC results to calculate the weighted average of the longitudinal and transverse components of the 
magnetization as an estimate of powder averaged neutron scattering data  in polycrystalline samples. This will be 
useful in analyzing experimental results in quantum magnets where large single crystals
are not available.

%{This paper deals with the magnetics of a two dimensional XXZ model when a magnetic field is applied transverse to the spin anisotropy direction. 
%The field causes a tilting of the sublattice magnetization producing canted magnetic orders in the system. For a small field, the overall magnetization 
%grows slowly as the field competes with the spin anisotropy. This also reduces
%the quantum fluctuation resulting in an unentangled ground state at a finite field value. Beyond this point, however, fluctuation shoots up fast to become maximum at the transition point $h_c$ and we need an analytic theory above mean-field level to describe such critical points. Thus our SWT$^{(2)}$ calculation gives
%much more reliable phase boundaries compared to that from a SWT$^{(1)}$ approach.}
%
%{Moreover, we consider cases of powder samples, as used in many Neutron diffraction results, where there is no defined transverse or longitudinal directions. The powder averages of the magnetization from a XXZ antiferromagnet are computed using a QMC simulation. We find that}
%in a longitudinal field, the saturation field can be calculated exactly, {$h_{s}=zS(J_{\perp}+J_{z})$}. In a transverse field,
%the expression is no longer exact, in part because the saturation field is replaced by a critical field. However, keeping terms up to first order 
%in perturbation theory the critical field becomes {$h_{c}=zS(3J_{\perp}+J_{z})/2$}.

\begin{acknowledgments}
{\em Acknowledgments.---}SK thanks K. Sengupta for useful discussions. Financial support from CSIR, India, under Scientists' Pool Scheme No. 13(8764-A)/2015-Pool (SK) and from the Ministry of Education, Singapore through Grant No. MOE2014-T2-1-112 (PS) are  gratefully acknowledged.
\end{acknowledgments}

%%%%%%%%%%%%%%%%%%%%%%%%%
% Begin the Appendix / Supplemental Material   %
%%%%%%%%%%%%%%%%%%%%%%%%%

\appendix

%\section{Supplemental material for ``Magnons in a two dimensional transverse field XXZ model''}

%%%%%%%%%%%%%%%%%%%%%
\section{Details of SWT$_a^{(1)}$}
%%%%%%%%%%%%%%%%%%%%%
\label{ap1}

When we write down the Hamiltonian for $\Delta<1$ in terms of the sublattice rotations, we obtain
%\vskip 3 in
\begin{eqnarray}
 H&=&\sum_{<ij>}H_{ij}=\sum_{<ij>}[\Delta S_i^zS_j^z+{\rm cos}(2\theta)(S_i^xS_j^x+S_i^yS_j^y)\nonumber\\
 &&+{\rm sin}(2\theta)(S_i^xS_j^y-S_i^yS_j^x)\nonumber\\
 &&-\frac{h}{Z}((S_i^x+S_j^x){\rm cos}\theta-(S_i^y-S_j^y){\rm sin}\theta)].
\end{eqnarray}
Here Hamiltonian is written in units of $J_{xy}$.
With $x$ being the quantization axis and $S_i^\pm=S_i^y\pm iS_i^z$ the raising and lowering operators, we can rewrite the Hamiltonian
as
\begin{eqnarray}
 H&=&\sum_{<ij>}[{\rm cos}(2\theta)S_i^xS_j^x-\frac{h}{Z}(S_i^x+S_j^x){\rm cos}\theta\nonumber\\
 &&+\frac{{\rm cos}2\theta}{4}(S_i^++S_i^-)(S_j^++S_j^-)\nonumber\\
 &&-\frac{\Delta}{4}(S_i^+-S_i^-)(S_j^+-S_j^-)\nonumber\\
 &&+{\rm sin}2\theta((S-n_i)\frac{S_j^++S_j^-}{2}-(S-n_j)\frac{S_i^++S_i^-}{2}))\nonumber\\
 &&+\frac{h}{Z}(\frac{S_i^++S_i^-}{2}-\frac{S_j^++S_j^-}{2}){\rm sin}\theta]
 \end{eqnarray}
 Now applying Holstein Primakoff transformation for SWT in a ferromagnet, we get $S_i^x=S-a_i^\dagger a_i=S-n_i$ 
 ($S_j^x=S-b_j^\dagger b_j=S-n_j$) and $S_i^+=\sqrt{2S}a_i$ ($S_j^+=\sqrt{2S}b_j$)  where $i$ ($j$) denotes the $\uparrow$ ($\downarrow$) 
 sublattice along $x$, and $a_i$'s ($b_j$'s) are the bosonic operators in the $\uparrow$ ($\downarrow$) sublattice.
 Hence we obtain,
 \begin{align}
 H&=\hspace{-.1 in}\sum_{<ij>}[{\rm cos}(2\theta)(S-n_i)(S-n_j)-\frac{h}{Z}(2S-n_i-n_j){\rm cos}\theta\nonumber\\
 &+\frac{{\rm cos}2\theta-\Delta}{4}(a_i^\dagger b_j^\dagger+{\rm hc})+\frac{{\rm cos}2\theta+\Delta}{4}(a_i^\dagger b_j+{\rm hc})\nonumber\\
 &+\frac{2h{\rm sin\theta/Z-sin}2\theta}{4}(a_i^\dagger-b_j^\dagger+{\rm hc})]\nonumber\\
&=\hspace{-.05 in}E_0(\theta)+\sum_{<ij>}[-cos(2\theta) S(n_i+n_j)+\frac{h}{Z}(n_i+n_j)cos\theta\nonumber\\
&+\frac{{\rm cos}2\theta-\Delta}{4}(a_i^\dagger b_j^\dagger+hc)+\frac{cos~2\theta+\Delta}{4}(a_i^\dagger b_j+ hc)\nonumber\\
&+\frac{2h{\rm sin\theta/Z-sin}2\theta}{4}(a_i^\dagger-b_j^\dagger+{\rm hc})] .
\end{align}
By minimizing $E_0(\theta)$, we obtain the reference angle $\theta_r$ as cos$\theta_r=h/2ZS$.
Thus we fix the reference state for spin wave expansion.
A consecutive Fourier transformation, thereafter, leads to
\begin{eqnarray}
H&=&E_0(\theta_r)+\sum_{k}[\frac{Z}{2}(a_k^\dagger a_k+b_k^\dagger b_k)+Z\gamma_k(\frac{{\rm cos}2\theta_r+\Delta}{4}\nonumber\\
&&(a_k^\dagger b_{k}+hc)+\frac{{\rm cos}2\theta_r-\Delta}{4}(a_k^\dagger b_{-k}^\dagger+hc))].
\label{ham-af}
\end{eqnarray}
Finally a Bogoliubov transformation brings in the magnon modes to be given by $\Omega_k=\sqrt{(\frac{Z}{2}\pm \frac{Z\gamma_k({\rm cos}2\theta_r+\Delta)}{4})^2-(\frac{Z\gamma_k({\rm cos}2\theta_r-\Delta)}{4})^2}$.
Notice that for $cos(2\theta_r)=\Delta$, the Hamiltonian \ref{ham-af} becomes diagonal making the reference state there the actual 
factorized ground state. So at the factorization point, cos$\theta_f=\sqrt{\frac{1+\Delta}{2}}$ and $h_f=2ZS$cos$\theta_f$.

Now for  $\Delta>1$, we will have
\begin{align}
 H&=\hspace{-.1 in}\sum_{<ij>}[S_i^xS_j^x(cos^2\theta-\Delta sin^2\theta)+\frac{(S_i^++S_i^-)(S_j^++S_j^-)}{4}\nonumber\\&
 +\frac{sin(2\theta)(1+\Delta)}{4i}(S_i^x(S_j^+-S_j^-)-(S_i^+-S_i^-)S_j^x)\nonumber\\
 &-\frac{(S_i^+-S_i^-)(S_j^+-S_j^-)}{4}(-sin^2\theta+\Delta cos^2\theta)\nonumber\\
 &-\frac{h}{Z}((S_i^x+S_j^x){\rm cos}\theta-\frac{(S_i^+-S_i^-)-(S_j^+-S_j^-)}{2i}{\rm sin}\theta)].\nonumber\\
\end{align}
within linear spin wave theory which becomes
\begin{align}
H&=E_0(\theta)+\sum_{<ij>}[(\Delta sin^2\theta-cos^2\theta) S(n_i+n_j)+\nonumber\\
&\frac{h}{Z}(n_i+n_j)cos\theta+\frac{1+sin^2\theta-\Delta cos^2\theta}{4}(a_i^\dagger b_j^\dagger+hc)\nonumber\\
&+\frac{1-sin^2\theta+\Delta cos^2\theta}{4}(a_i^\dagger b_j+ hc)+\nonumber\\
&(\frac{hsin\theta}{2Zi}-\frac{sin(2\theta)(1+\Delta)}{4i})((a_i^\dagger-b_j^\dagger-hc)] .\nonumber
\end{align}
Minimizing $E_0(\theta)$ gives, $cos\theta_r=h/SZ(1+\Delta)$. And with this and by Fourier transformation we obtain
\begin{align}
H&=E_0(\theta_r)+Z\sum_{k}[\frac{\Delta}{2}(a_k^\dagger a_k+b_k^\dagger b_k)+\gamma_k(\frac{{\rm cos}^2\theta_r(1+\Delta)}{4}\nonumber\\
&(a_k^\dagger b_{k}+hc)+\frac{2-{\rm cos}^2\theta_r(1+\Delta)}{4}(a_k^\dagger b_{-k}^\dagger+hc))].
\label{ham-af2}
\end{align}
with  $cos\theta_r=h/SZ(1+\Delta)$. The factorizing point is denoted by  cos$\theta_f=\sqrt{\frac{2}{1+\Delta}}$ and $h_f=ZS(1+\Delta)$cos$\theta_f$.
The magnon modes are given by 
$\Omega_k=\sqrt{(\frac{\Delta Z}{2}\pm
\frac{Z\gamma_k{\rm cos}^2\theta_r(1+\Delta)}{4})^2-(\frac{Z\gamma_k(2-{\rm cos}^2\theta_r(1+\Delta))}{4})^2}$.

%%%%%%%%%%%%%%%%%%%%%%%%%%%%%%%%%%%%%
\section{Obtaining magnon modes from a $4\times4$ SW Hamiltonian}
%%%%%%%%%%%%%%%%%%%%%%%%%%%%%%%%%%%%%
\label{ap2}

Let's now construct the magnon modes from the $k$-space Hamiltonian,
\begin{equation}
 H=E_0+\sum_k[A_k(a_k^\dagger a_k+b_k^\dagger b_k)+(B_ka_k^\dagger b_k+C_ka_k^\dagger b_{-k}^\dagger +hc)].
\end{equation}
We can write this as $H=E_0+\sum_k <\phi_k|H_k|\phi_k>$ where $|\phi_k>=(a_k,b_{-k}^\dagger,b_k,a_{-k}^\dagger)^T$
and 
\begin{displaymath}
H_k=\left(\begin{array}{cccccccc}
A_k & C_k & B_k & 0\\
C_k & A_k & 0 & B_k \\
B_k & 0 & A_k & C_k\\
0 & B_k & C_k & A_k~~~
\end{array}\right).
\end{displaymath}
From there we can obtain the diagonalized version as outlined in Refs.~\citenum{mano} and~\citenum{kar}.
A Bogoliubov transformation brings in the states $|\psi_k>=U|\phi_k>$ where $|\psi_k>=(\alpha_k,\beta_{-k}^\dagger,\beta_k,\alpha_{-k}^\dagger)^T$ 
and $U$ a coefficient matrix so that $U^\dagger H_kU$ becomes a diagonal matrix with eigenvalues $\lambda_k$'s.
This as well as the bosonization of the new variables $\alpha_k$ and $\beta_k$ requires Det[$M_k$]=0 for a certain matrix $M_k$, given as 
\begin{displaymath}
M_k=\left(\begin{array}{cccccccc}
A_k-\lambda_k & C_k & B_k & 0\\
C_k & A_k+\lambda_k & 0 & B_k \\
B_k & 0 & A_k-\lambda_k & C_k\\
0 & B_k & C_k & A_k+\lambda_k~~~
\end{array}\right).
\end{displaymath}
Hence we get $\lambda_{k(\pm)}=[(A_k\pm B_k)^2-C_k^2]^{0.5}$ and the Hamiltonian becomes
\begin{equation}
 H=E_0'+\sum_k[\lambda_{k(+)}\alpha_k^\dagger \alpha_k +\lambda_{k(-)}\beta_k^\dagger \beta_k].
\end{equation}
Solving for the coefficient matrix $U$ (see Ref.~\citenum{kar}), we can also obtain the spin deviation given as 
$\epsilon=\frac{1}{N}\sum_k(<a_k^\dagger a_k>+<b_k^\dagger b_k>)$, where $<..>$ denotes the ground state average.

%%%%%%%%%%%%%%%%%%%%%
\section{Details of SWT$_b^{(1)}$}
\label{ap3}
%%%%%%%%%%%%%%%%%%%%%

On the other hand, if we want to do the spin wave analysis for large $h$ values we rather consider the ferromagnetic spin orientations
along the field direction $x$ to be the quantization axis and take that ferromagnetic state (with no sublattice division) to be the spin reference state.
A $\pi/2$ rotation about the $y$ axis moves the $z$ axis to the field direction and that becomes the $z$ axis in the transformed coordinates. 
Within such definition, the transverse field XXZ Hamiltonian becomes
\begin{eqnarray}
H&=&\sum_{\left<ij\right>}[S_{i}^{z}S_{j}^{z}+
\frac{\Delta-1}{4}\left(S_{i}^{+}S_{j}^{+}+hc\right)+\nonumber\\
&&\frac{1+\Delta}{4}\left(S_{i}^{+}S_{j}^{-}+hc\right)]-h\sum_{i}S_{i}^{z}.
\end{eqnarray}
Using linear spin wave theory, this becomes,
\begin{eqnarray}
H&=&E_0+(h-SZ)\sum_in_i+\sum_{\left<ij\right>}(\frac{1+\Delta}{4}a_i^\dagger a_j\nonumber\\
&&+\frac{\Delta-1}{4}a_i^\dagger a_j^\dagger +hc).\nonumber\\
\end{eqnarray}
Then we do the Fourier transformation to get
\begin{eqnarray}
H&=&E_0+(h-SZ)\sum_kn_k+\sum_k[\frac{1+\Delta}{2}\gamma_kSZa_k^\dagger a_k\nonumber\\
&&+\frac{\Delta-1}{2}\gamma_kSZ(a_ka_{-k}+hc)]
\end{eqnarray}
Finally a Bogoliubov transformation gives
\begin{eqnarray}
 H=E_0'+\sum_k\Omega_k{\tilde a}_k^\dagger {\tilde a}_k,
\end{eqnarray}
where $\Omega_k=\sqrt{(h-SZ+\frac{1+\Delta}{2}\gamma_kSZ)^2-(\frac{\Delta-1}{2}\gamma_kSZ)^2}$.
So this gives the critical $h$ to be ${h_c'}=SZ+\frac{1+\Delta}{2}SZ+\frac{|\Delta-1|}{2}SZ$.

%%%%%%%%%%%%%%%%%%%%%%%%%%%%%
\section{Second order Spin wave theory at large fields}
%%%%%%%%%%%%%%%%%%%%%%%%%%%%%
\label{ap4}

In order to do a second order correction to the linear SWT$_b^{(1)}$ at high fields,
we see that the higher order correction to spin wave expansion, for a 
ferromagnetic reference state, gives us a modified $S_i^+=\sqrt{2S}(a_i-\frac{n_ia_i}{2})$.
This alters the off-site interaction terms as
\begin{eqnarray}
S_i^+S_j^-&=&a_ia_j^\dagger(1-\frac{n_j}{2})-\frac{n_ia_ia_j^\dagger}{2}\nonumber\\
S_i^+S_j^+&=&a_ia_j(1-\frac{n_j}{2})-\frac{n_ia_ia_j}{2}
\end{eqnarray}
up to the quartic order of the bosonic operators.
A mean field treatment for a product of variables $A$ and $B$ can be given as $AB=A\overline{B}+\overline{A}B$
where $\overline{A}$ and $\overline{B}$ are the respective averages. Applying that 
to the quartic correction terms, we obtain the  modified the magnon mode expressions of
SWT$_b^{(1)}$ to be 
\begin{align}
\Omega_k&=\sqrt{A_{1k}^2-A_{2k}^2}~~~~{\rm where}\nonumber\\
&A_{1k}=h-SZ[1-\frac{1+\Delta}{2}\{\gamma_k(1-\epsilon)-\delta\}+(\Delta-1)\eta]\nonumber\\&
{\rm and}~~~ A_{2k}=\frac{\Delta-1}{2}\gamma_kSZ(1-\epsilon). 
\end{align}
The critical field becomes 
\begin{eqnarray}
{h_c^{(2)}}&=SZ[1+\frac{1+\Delta}{2}(1-\epsilon-\delta)-(\Delta-1)\eta\nonumber\\&~~~~~+\frac{|\Delta-1|}{2}(1-\epsilon)]
\end{eqnarray}
Here $<\overline{n_i}>=<\overline{n_j}>=\epsilon$ is the spin deviation. The other fluctuation measures 
$\delta=<\overline{a_i^\dagger a_j}>=\sum_k\gamma_ka_k^\dagger a_k/N$ and
$\eta=<\overline{a_i a_j}>=\sum_k\gamma_ka_k a_{-k}/N$ (see Ref.\onlinecite{mila2}).
So we need to calculate both $\Omega_k$ and $h_c$ numerically in a 
self-consistent manner. We obtain the critical fields at Ising and XY limit to be ${h_c^{(2)}|_{Ising}}=1.75$ and ${h_c^{(2)}|_{XY}}=3.75$.
See that for the isotropic point $\Delta=1$, $h_c'=2Zs=4$ is the factorization point where $\epsilon=0$. Other fluctuations
$\delta$ and $\eta$ are also zero at this point, which have been checked numerically.

%%%%%%%%%%%%%%%%%%%%%%%
\section{Details of Quantum Monte Carlo}
%%%%%%%%%%%%%%%%%%%%%%%
\label{ap5}
%%%%%%%%%%%%%%%%%%%%%%%%%%%%
% FIGURE 7: vertices for transverse field XXZ model  %
%%%%%%%%%%%%%%%%%%%%%%%%%%%%

\begin{figure}[t]
\centering
\includegraphics[width=0.85\linewidth]{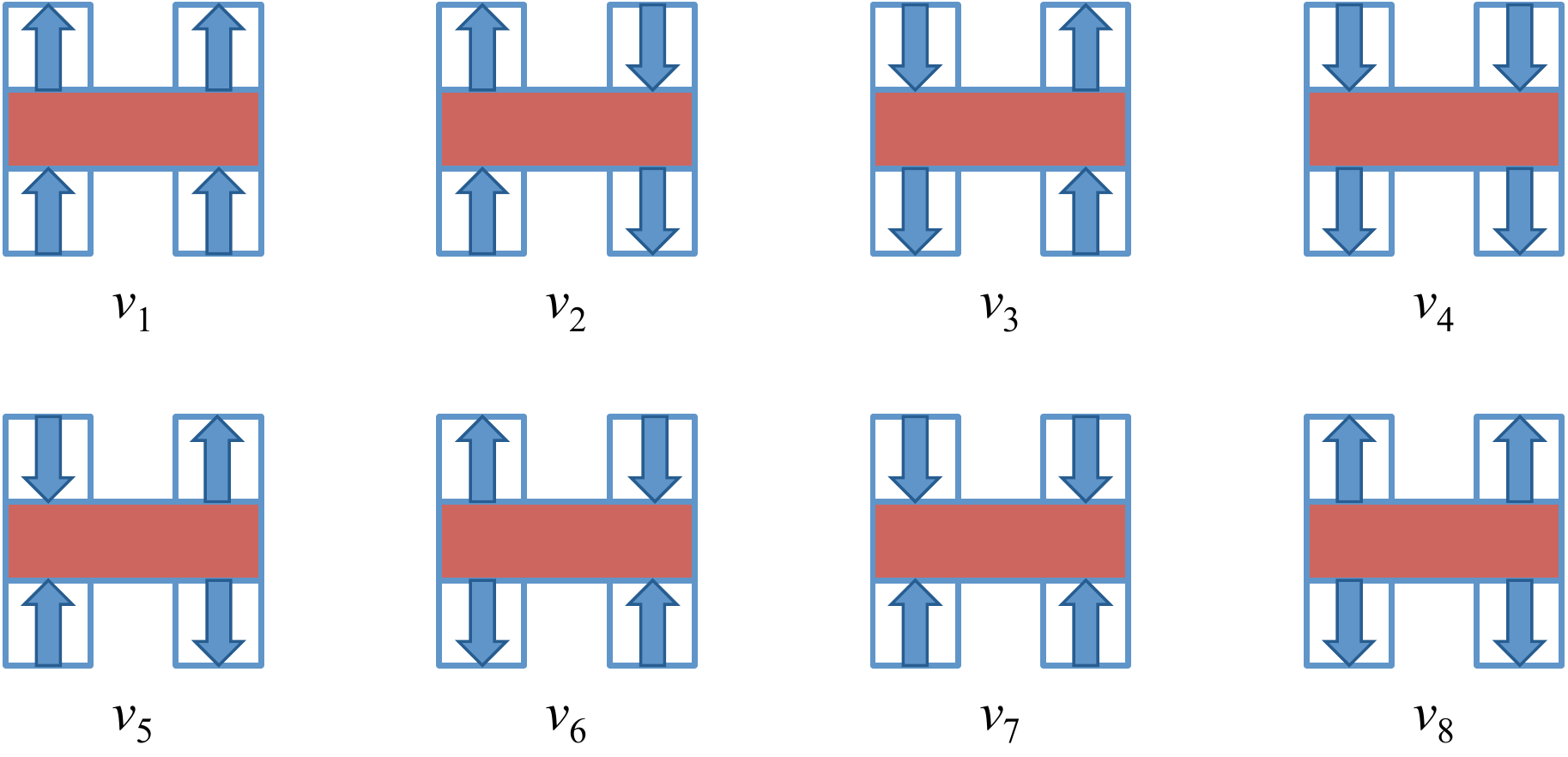}
\caption{(Color online) The allowed diagonal ($v_{1}$--$v_{4}$) and off-diagonal ($v_{5}$--$v_{6}$) vertices for the TF-XXZ model. The first six vertices also appear in the longitudinal field XXZ model, while the last two vertices appear whenever $\Delta\neq1$.}
\label{fig-vertex}
\end{figure}

In Fig.~\ref{fig-vertex} we show the allowed vertices for the TF-XXZ model. This includes the addition of two number-nonconserving vertices ($v_{7}$ and $v_{8}$) to the usual six vertices ($v_{1}$--$v_{6}$) of the XXZ model.
On a bipartite lattice, it can be shown that the vertices $v_{5}$ and $v_{6}$ must occur an even number of times ($n_{v_{5}}+n_{v_{6}}$ is even), which is sufficient to ensure that the overall contribution to the weight function in the {\em diagonal sector} is positive definite. Similarly, the vertices $v_{7}$ and $v_{8}$ must also occur an even number of times ($n_{v_{7}}+n_{v_{8}}$ is even), even on non-bipartite lattices. 

Measurements in the {\em off-diagonal sector} are also possible, but the total weight is no longer guaranteed to be positive definite. However, since the partition function is still defined in the diagonal sector, the total weight in the off-diagonal sector can be obtained by working with the absolute weights while keeping track of the overall sign of the vertex weights. Observables such as the Green's function are then calculated as the signed average over configurations. In short, measurements in the off-diagonal sector are easily obtained by using the absolute value of all off-diagonal vertex weights, while separately keeping track of the overall sign of the operator string as it evolves during the loop update. We find this method to be much simpler in practice than the standard alternative: first define a formal (sub)lattice transformation such that all off-diagonal terms become negative definite, then determine the momentum shift required to map between the original and transformed Hamiltonian observables. For the XXZ model in a longitudinal field, this becomes a sublattice rotation of $\pi$ around the $z$ axis, with a $(\pi,\pi)$ momentum shift. In the present case, an additional lattice rotation by $\pi/2$ is required whenever $\Delta>1$, which becomes tedious to keep track of compared to the relative simplicity of our explicit sign-tracking described above. Another benefit to our method of sign tracking is that if the overall sign is ever negative at the close of the loop update, then we know that the model has a QMC sign problem. Thus, we have explicitly checked our assumption that no sign problem exists for the TF-XXZ model as defined in this paper.

%%%%%%%%%%%%%%%%%%%%%%
\section{Details of Perturbation Theory}
%%%%%%%%%%%%%%%%%%%%%%
\label{ap6}

Let us begin by writing the unperturbed Hamiltonian as
\begin{eqnarray}
\begin{split}
{\cal H}_{0}=J_{\perp}\sum_{\langle ij\rangle}S_{i}^{x}S_{j}^{x}-h\sum_{i}S_{i}^{x},
\end{split}
\end{eqnarray}
so that the perturbed Hamiltonian becomes
\begin{eqnarray}
\begin{split}
{\cal H}^{'}=J_{\perp}\sum_{\langle ij\rangle}
\frac{1-\Delta}{4}\left(S_{i}^{+}S_{j}^{+}+S_{i}^{-}S_{j}^{-}\right)+\\
\frac{1+\Delta}{4}\left(S_{i}^{+}S_{j}^{-}+S_{i}^{-}S_{j}^{+}\right).
\end{split}
\end{eqnarray}
Next, we consider the zeroth-order (unperturbed) contribution to the energy of a state with all and all-but-one of its spins aligned with the magnetic field, and label these energies $E_{0}(N\uparrow,0\downarrow)$ and $E_{0}(N-1\uparrow,1\downarrow)$, respectively. It is easy to show that
\begin{eqnarray}
\begin{split}
E_{0}(N\uparrow,0\downarrow)&=\frac{NZ}{2}J_{\perp}S^{2} - NBS \\
E_{0}(N-1\uparrow,1\downarrow)&=\left(\frac{NZ}{2}-2Z\right)J_{\perp}S^{2} - \left(N-2\right)BS.
\end{split}
\end{eqnarray}
The first-order corrections can be obtained as $\langle\psi_{0}|{\cal H}^{`}|\psi_{0}\rangle$, and are given by
\begin{eqnarray}
\begin{split}
E_{1}(N\uparrow,0\downarrow)&=0 \\
E_{1}(N-1\uparrow,1\downarrow)&=-Z\frac{J_{\perp}+J_{z}}{4}.
\end{split}
\end{eqnarray}
Finally, by equating these energies up to first order (i.e. $E_{0}+E_{1}$) we find an estimate of the critical field, where the fully saturated unperturbed state is favorable to the state with a flipped spin: $h_{c}=Z(3J_{\perp}+J_{z})/4$.


\begin{thebibliography}{99}

\bibitem{sub-sach} {S. Sachdev, Nature Physics {\bf 4}, 173 (2008).}

\bibitem{bec} V. Zapf, M. Jaime, and C. D. Batista, Rev. Mod. Phys. {\bf 86}, 563 (2014).

\bibitem{spinliquids} L. Balents, Nature {\bf 464}, 199 (2010).

\bibitem{vbs} {A. W. Sandvik, Phys. Rev. Lett. {\bf 98}, 227202 (2007).}

\bibitem{vbs2} K. Matan $et~al.$, Nature Physics {\bf 6}, 865 (2010).

\bibitem{owerre} {S. A. Owerre, arXiv:1609.03563; arXiv:1701.05199 (unpublished).}

\bibitem{zhou2} J. Zhou $et~al.$, Phys. Rev. Lett. {\bf 116}, 256601 (2016).

\bibitem{mag-plat} {H. Kageyama $et~al.$ Phys. Rev. Lett. {\bf 82}, 3168 (1999).}



 \bibitem{Breunig2013} O. Breunig, M. Garst, E. Sela, B. Buldmann, P. Becker, L. Bohat\'{y}, R. M\"{u}ller, and T. Lorenz, Phys. Rev. Lett. {\bf 111}, 187202 (2013).

  \bibitem{breunig2} O. Breunig, M. Garst, E. Sela, B. Buldmann, P. Becker, L. Bohat\'{y}, R. M\"{u}ller, and T. Lorenz, Phys. Rev. B {\bf 91}, 024423 (2015).

  \bibitem{kenzel} O. Kenzelmann, M. Garst, E. Sela, B. Buldmann, P. Becker, L. Bohat\'{y}, R. M\"{u}ller, and T. Lorenz, Phys. Rev. B {\bf 65}, 144432 (2002).

\bibitem{Cuccoli2003} {A. Cuccoli, T. Roscilde, V. Tognetti, R. Vaia, and P. Verrucchi, Phys. Rev. B {\bf 67}, 104414 (2003).}

\bibitem{Holtschneider2005} {M. Holtschneider, W. Selke, and R. Leidl, Phys. Rev. B {\bf 72}, 064443 (2005).}

\bibitem{yunoki} {S. Yunoki, Phys. Rev. B {\bf 65}, 092402 (2002).}

\bibitem{Jensen2006} P. J. Jensen, K. H. Bennemann, D. K. Morr, and H. Dreyss\'{e}, Phys. Rev. B {\bf 73}, 144405 (2006).

\bibitem{Roscilde200x} T. Roscilde, P. Verrucchi, A. Fubini, S. Haas, and V. Tognetti, Phys. Rev. Lett. {\bf 93}, 167203 (2004); {\bf 94}, 147208 (2005).

\bibitem{langari} J. Abouie, A. Langari, M. Siahatgar, J. Phys.: Cond. Mat. {\bf 22}, 216008 (2010).



\bibitem{mahdavifar} H. Moradmard, M. Shahri Naseri, S. Mahdavifar, J. Supercond. Nov. Magn. {\bf 27}, 1265 (2014).
  
\bibitem{amico} L. Amico, R. Fazio, A. Osterloh, and V. Vedral, Rev. Mod. Phys. {\bf 80}, 517 (2008).

\bibitem{kurmann} J. Kurmann, H. Thomas, and G. Muller, Physica A {\bf 112}, 235 (1982).

\bibitem{amico2} L. Amico, F. Baroni, A. Fubini, D. Patan\`{e}, V. Tognetti, and P. Verrucchi, Phys. Rev. A {\bf 74}, 022322 (2006).

\bibitem{Brambleby2015} J. Brambleby, P. A. Goddard, R. D. Johnson, J. Liu, D. Kaminski, A. Ardavan, A. J. Steele, S. J. Blundell, T. Lancaster, P. Manuel, P. J. Baker, J. Singleton, S. G. Schwalbe, P. M. Spurgeon, H. E. Tran, P. K. Peterson, J. F. Corbey, and J. L. Manson, Phys. Rev. B {\bf 92}, 134406 (2015).


\bibitem{Oshikawa1997} M. Oshikawa and I. Affleck, Phys. Rev. Lett. {\bf 79}, 2883 (1997).

\bibitem{Essler1999} F. H. L. E\ss ler, Phys. Rev. B {\bf 59}, 14376 (1999).

\bibitem{Zvyagin2011} S. A. Zvyagin, E. \v{C}i\v{z}m\'{a}r, M. Ozerov, J. Wosnitza, R. Feyerherm, S. R. Manmana, and F. Mila, Phys. Rev. B {\bf 83}, 060409 (2011).


\bibitem{mano} E. Manousakis, Phys. Rev. B {\bf 79}, 220509 (2009).

\bibitem{kar} S. Kar, JMMM {\bf 393}, 357 (2015).

\bibitem{mila} L. P. Henry, P. C. W. Holdsworth, F. Mila, and T. Roscilde, Phys. Rev. B {\bf 85}, 134427 (2012).

%\bibitem{Wang2015} Z. Wang, J. Wu, S. Xu, W. Yang, C. Wu, A. K. Bera, A. T. M. Nazmul Islam, B. Lake, D. Kamenskyi, P. Gogoi, H. Engelkamp, A. Loidl, and J. Deisenhofer, arXiv:1512.01753v1 (2015).
{\bibitem{tsvelik} A. A. Nersesyan and A. M. Tsvelik, Phys. Rev. B {\bf 67}, 024422 (2003).}

\bibitem{Henelius200x} P. Henelius, A. W. Sandvik, C. Timm, and S. M. Girvin, Phys. Rev. B {\bf 61}, 364 (2000); P. Henelius, P. Fr\"{o}brich, P. J. Kuntz, C. Timm, and P. J. Jensen, {\it ibid.} {\bf 66}, 094407 (2002).

\bibitem{Syljuasen2003} O. F. Sylju\r{a}sen, Phys. Rev. E {\bf 67}, 046701 (2003).

\bibitem{Wenzel2008} S. Wenzel and W. Janke, Phys. Rev. B {\bf 78}, 064402 (2008).

\bibitem{Wenzel2010} S. Wenzel, W. Janke, and A. M. L\"{a}uchli, Phys. Rev. E {\bf 81}, 066702 (2010).


\bibitem{pelissetto} Pelissetto, Vicari, Phys. Rep. {\bf 368}, 549 (2002).

\bibitem{Rigol2007} M. Rigol and R. R. P. Singh, Phys. Rev. B {\bf 76}, 184403 (2007).

\bibitem{mila2} T. Coletta, N. Laflorencie, and F. Mila, Phys. Rev. B {\bf 85}, 104421 (2012).


\end{thebibliography}
\end{document}